\documentstyle[aps,epsf]{revtex}
\begin{document}
\newcommand{\chem}[1]{${\rm #1}$}
\newcommand{\half}[0]{$\frac{1}{2}$}
\title{Interplay of Chemical Bonding and Magnetism in \chem{Fe_4N},
\chem{Fe_3N} and \chem{\zeta - Fe_2N}}
\author{M. Sifkovits, H. Smolinski, S. Hellwig, and W. Weber}
\address{Institute of Physics, University of Dortmund, Germany}
\maketitle
\begin{abstract}
Using spin density functional theory we have carried out a comparative study 
of chemical bonding and magnetism in \chem{Fe_4N}, \chem{Fe_3N} and 
\chem{\zeta - Fe_2N}. All of these compounds form close packed Fe lattices, while N
occupies octahedral interstitial positions. High spin fcc Fe and
hypothetical FeN with rock salt structure have been included in our study as 
reference systems. We find strong, covalent Fe--N bonds as a result of a 
substantial $\sigma$--type p--d hybridisation, with some charge transfer to
N. Those Fe d orbitals which contribute to the p--d bonds, do no longer 
participate in the exchange splitting of the Fe d bands. Because of the
large exchange fields, the majority spin d bands are always fully occupied, 
while the minority spin d bands are close to half--filling, thus optimizing
the Fe d--d covalent bonding. As a consequence, in good approximation
the individual Fe moments decrease in steps of $\frac{1}{2}$ \chem{\mu_B} 
from fcc
iron (2.7 \chem{\mu_B}) via \chem{Fe_4N} (2.7 and 1.97 \chem{\mu_B}), 
\chem{Fe_3N} (1.99 \chem{\mu_B}) to \chem{\zeta - Fe_2N}
(1.43 \chem{\mu_B}). \\
\end{abstract}


{\bf \noindent PAC's : } \\
75.50.-y; 75.50.Bb; 75.50.Ss\\

{\bf \noindent Postal Adress : } \\
Prof. Dr. Werner Weber \\
Universit{\"a}t Dortmund, Institut f{\"u}r Physik \\
Lehrstuhl f{\"u}r theoretische Physik II \\
44221 Dortmund \\
FAX : ++49 0231 755 3569 \\
Tel : ++49 0231 755 3563 \\ 
E-mail : weber@fkt.physik.uni-dortmund.de

\section{Introduction}

The nitrides of Fe have since long played an important role in steel 
technology. In particular, nitriding is used to harden iron surfaces and to 
passivate them against oxidation. On the other hand, the magnetic properties of
some nitrides, especially those with small N contents, have raised considerable
interest concerning magnetic data storage applications. \\
The phase diagram of \chem{FeN_x} has been studied by the classic work of 
Jack \cite{Jack1,Jack2}. Apart from the regions of very small N concentrations,
 such as \chem{Fe_{16}N_2} (derived from the bcc iron), most ordered nitrides of Fe 
are based on close packed Fe structures. \chem{Fe_4N} may be seen as fcc Fe, 
where $\frac{1}{4}$ of all octahedral sites are occupied by N in such a way,
that the octahedra share corners only. The structure of \chem{Fe_3N} is based
on a hexagonal close packed lattice of Fe, now with $\frac{1}{3}$ of the 
octahedral sites being filled by N. Again, these are corner sharing octahedral
units only. Finally, the \chem{\zeta - Fe_2N} lattice is also based on hcp Fe, and 
$\frac{1}{2}$ of the octahedral sites are occupied by N; now edge sharing 
arrangements appear as well. \\
In recent years, details of the \chem{FeN_x} phase diagram, especially in the
range between \chem{FeN_{0.3}} and \chem{FeN_{0.5}} have been investigated by 
Rechenbach et al. \cite{Rechenbach1,Rechenbach2}. This study also includes neutron scattering
data, which allow to extract the spatial distribution of the magnetic moment
density.\\
Electronic structure calculations of the ordered iron nitrides based on spin 
density functional theory (SDFT) have been reported by various groups
\cite{Matar1,Matar2,Matar3,Sakuma1,Sakuma2}. 

We employ the highly accurate full potential LAPW method of the WIEN95 program
package\cite{Wien95}. For a systematic determination of ionic charges, we use
the zero--flux charge density gradient method of Bader et al. \cite{Bader}.
In order to follow the trends in bonding and in magnetic properties through the
series of nitrides, we include as references results for fcc Fe in the high
spin state and also results for FeN in rock salt structure. \\

\section{Calculational procedures}

The structural parameters of the three nitrides were taken from experiment
\cite{Rechenbach1,Rechenbach2}.
They are listed in table \ref{structures}. For our SDFT electronic structure 
calculations we used the local density approximation (LDA) and employed the 
exchange--correlation potential proposed by Perdew and Wang \cite{Perdew}.
In all our calculations we used the same muffin--tin radii 
\chem{R_{MT}(Fe)} = 1.7 and \chem{R_{MT}(N)} = 1.5 atomic units. Typical 
energy cut--offs were between \chem{R_{MT}(N)*K_{max}} = 8 and 10. This 
corresponds to more than 150 plane waves per atom. For \chem{Fe_4N} we used a 
k--mesh
with 120 points in the irreducible part of the Brillouin zone, for 
\chem{Fe_3N} and \chem{\zeta - Fe_2N} the corresponding numbers have been 116 and 
100 points. To accelerate 
self--consistency, we used a Fermi--factor smearing temperature \chem{T_{FFS}}
= 100 K. \\
For the determination of ionic charges by the zero--flux gradient method, a 
very fine real space mesh of the charge density \chem{\rho(\vec{r})} was 
required. In fact each basis vector of the unit cell was divided into up to
400 intervals. \\

\section{Results}

\subsection{High spin fcc Fe}

\begin{figure}
\epsfysize=13cm
\epsffile{ 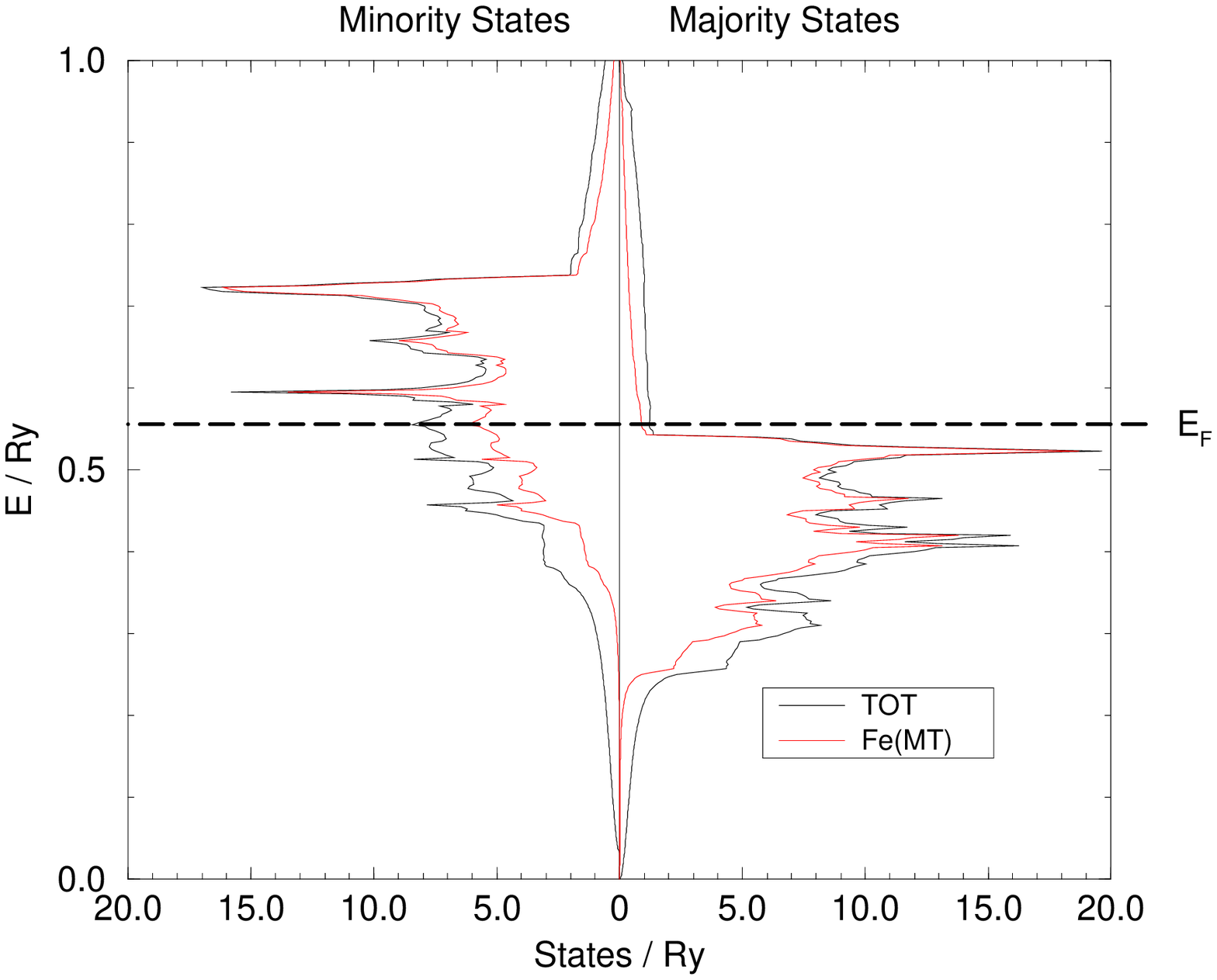}
\caption{\label{dos_fccfe} Total density of states for high spin fcc Fe.
}
\end{figure}

\begin{figure}
\epsfysize=8 cm
\hspace{1.5cm}
\epsffile{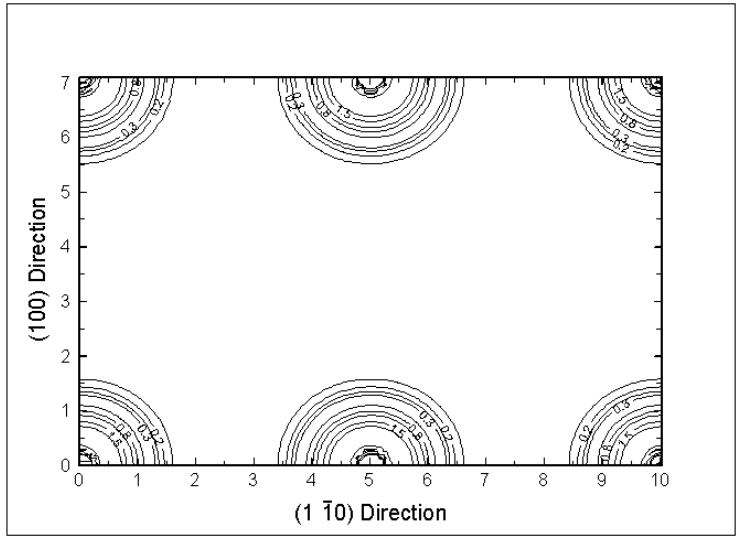}
\caption{\label{charge_fccfe} Valence charge density of fcc Fe in the 
(1$\bar{1}$0) plane in units of e/(atomic units $)^3$. The length scales are
also in a.u. }
\end{figure}

\begin{figure}
\epsfysize=8 cm
\hspace{1.5cm}
\epsffile{ 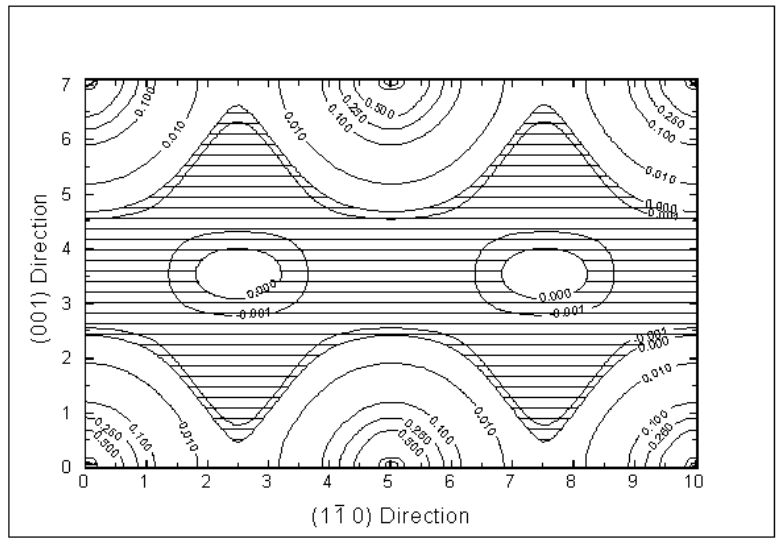}
\caption{\label{spin_fccfe} Spin density of fcc Fe in the (1$\bar{1}$0)
 plane in units of e/\chem{(a.u.)^3}. Areas of negative--spin density are 
hatched.}
\end{figure}

As a reference, we discuss the electronic structure of high spin fcc 
Fe (see Fig. \ref{dos_fccfe}) first. 
For the SDFT calculation, the lattice constant of \chem{Fe_4N} was used.
The majority spin d--bands are 
completely filled, while for the minority bands the Fermi energy lies close to
the center of the d--bands in a local minimum of the density of states. Evaluating the exchange splitting of some typical d--band states we find 
${\rm \delta^{AVG}_{E_g} = 2.7 }$ eV. The magnetic moment is 2.65 ${\rm \mu_B}$. The s--band, which
does not exhibit a large exchange splitting, is occupied by $ \approx $ 0.63 
electrons. The charge density (see Fig. \ref{charge_fccfe}) is rather 
spherical 
around the Fe atoms, yet there is some extra charge along the first nearest 
neighbour (1NN) Fe--Fe bond. This bonding charge is caused by the states of the
half filled minority bands (where all bonding states are occupied). Note that
the width of the minority d--bands is $\approx$ 1 eV larger than that of the
(fully occupied) majority d--bands. This band widening, typical of all 
half--filled d--band materials (see, e.g. \cite{Moruzzi}), is another 
consequence of 
covalent d--d bonding in the minority bands. It is caused by the larger 
radial extent of the d wave functions at half band filling. In comparison, the
 majority d--band wave functions are 
somewhat contracted radially, as the anti--bonding states are filled as well. 
As a 
consequence the minority band wave functions extend further into the 
interstitial regions, resulting in the negative--spin densities there (see
Fig. \ref{spin_fccfe}). 

\subsection{\chem{Fe_4N}}

\begin{figure}
\epsfysize=13cm
\epsffile{ 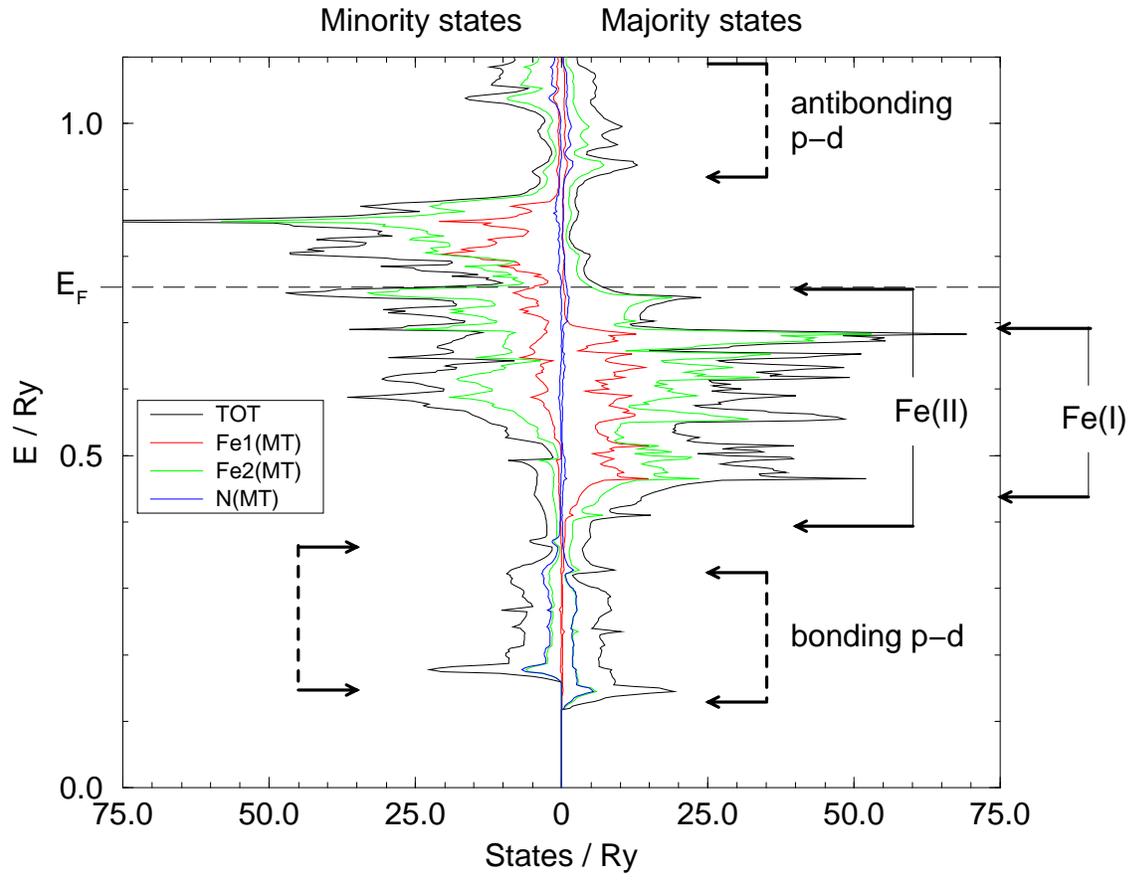}
\caption{\label{dos_fe4n} Total (TOT) and muffin--tin projected (MT)
densities
of states for \chem{Fe_4N}. Colours indicate MT orbital partial densities.}
\end{figure}

\begin{figure}
\epsfysize=8 cm
\hspace{1.5cm}
\epsffile{ 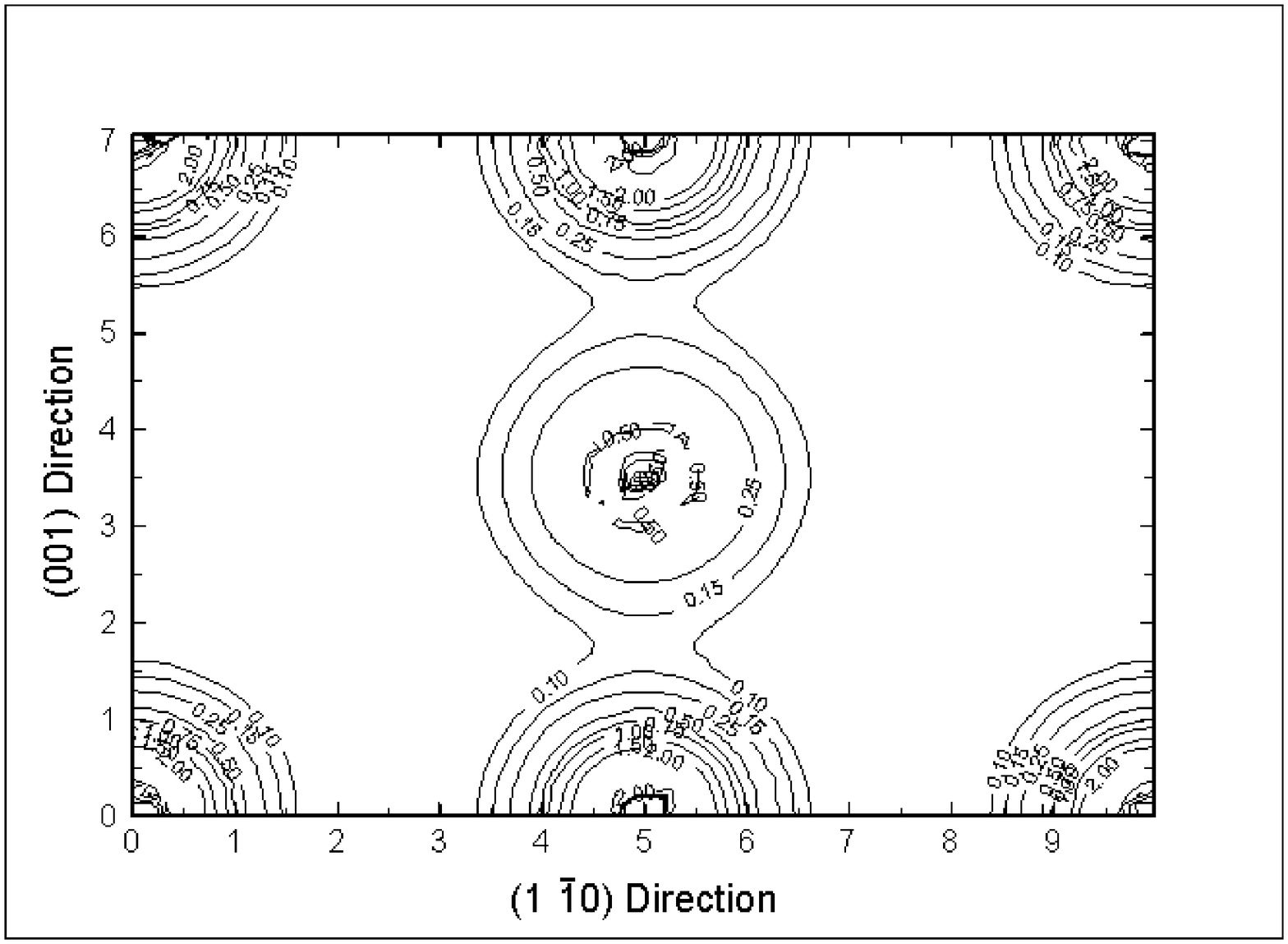}
\caption{\label{charge_fe4n} Valence charge density of \chem{Fe_4N} in the 
(1$\bar{1}$0) plane. }
\end{figure}

\begin{figure}
\epsfysize=8 cm
\hspace{1.5cm}
\epsffile{ 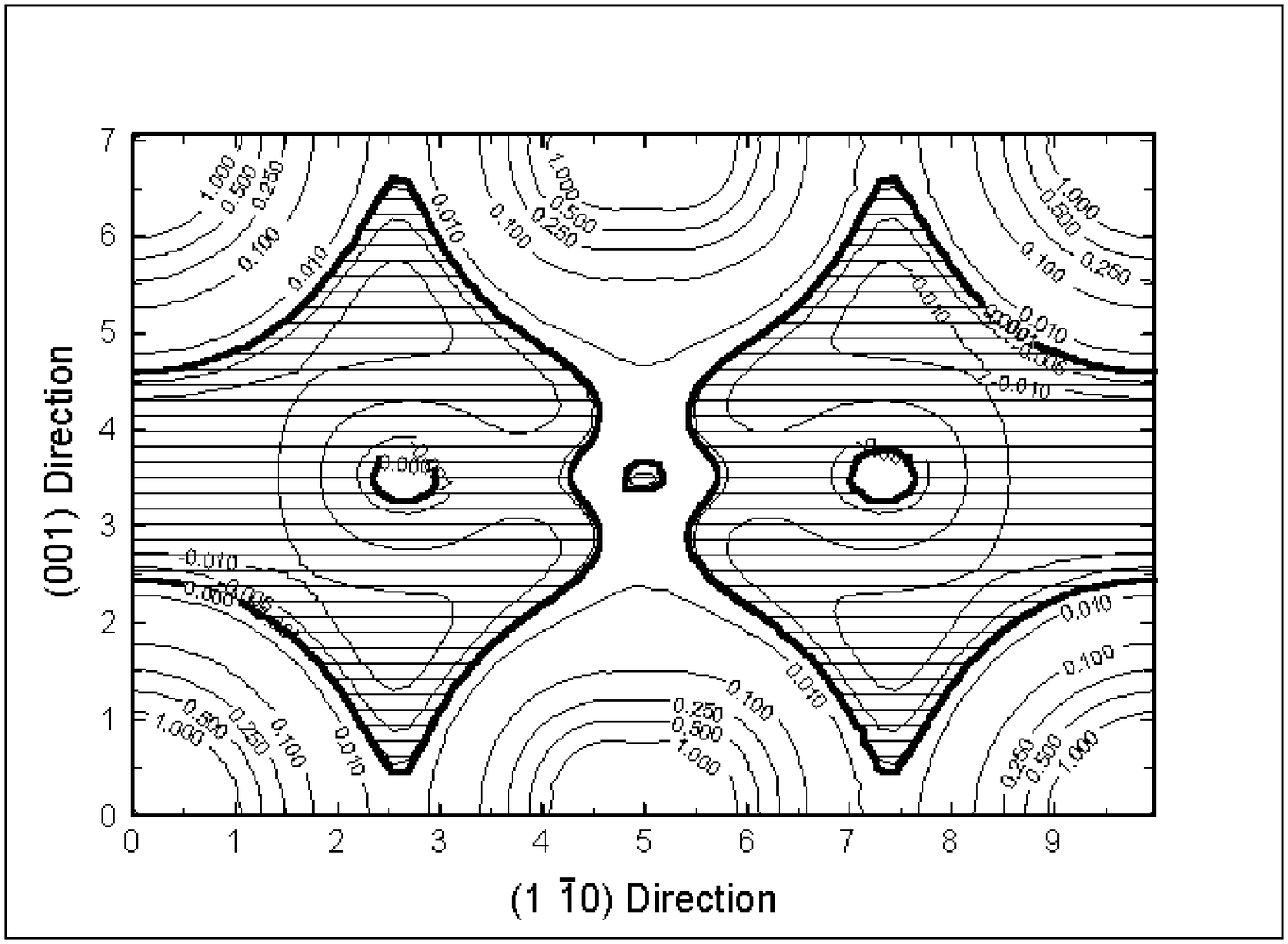}
\caption{\label{spin_fe4n} Spin density of \chem{Fe_4N} in the 
(1$\bar{1}$0) plane.}
\end{figure}

Only $\frac{1}{4}$ of the octahedral sites are occupied by  
nitrogen atoms, and 
there are two inequivalent iron sites. If we put N 
on the $(\frac{1}{2},\frac{1}{2},\frac{1}{2})$ position, there is one type of iron at (0,0,0) labeled Fe(I),  and the
other, Fe(II), at $(\frac{1}{2},\frac{1}{2},0)$, 
$(\frac{1}{2},0,\frac{1}{2})$, and $(0,\frac{1}{2},\frac{1}{2})$. \\
In the density of states curves in Fig. \ref{dos_fe4n} we 
can distinguish three band regions for both majority and minority spin bands. 
The
lowest one is made up by the N 2p states which are rather strongly hybridized
with the \chem{d(3z^2-r^2)} orbitals of the Fe(II) atoms through a 
\chem{(pd\sigma)} type coupling. Then there follow d--bands, which show hardly
any hybridisation with the N p orbitals. All of them are occupied in the 
majority spin bands. Above these d--bands there is the anti--bonding part of the 
p--d bands. Again, the Fermi level is situated close to the center of the 
d--band part of the minority spin. \\
The plot of the \chem{Fe_4N} valence charge density(see Fig. \ref{charge_fe4n})
indicates the strong covalent Fe(II)--N bond. In comparison the bonding charge
between Fe(I) and Fe(II) is by far less important. In a qualitative manner we 
may thus realize the extra stability provided by the 
addition of N as compared to the pure Fe lattice. \\
The spin density plot (Fig. \ref{spin_fe4n}) indicates that the dominant 
contributions to the spin density are located within the muffin--tin spheres.
There are  
areas of negative moment density in the interstitial regions, which is again 
caused by the larger extension of the minority--spin wave functions. However, 
the
total negative moment of those regions is rather small.  
The magnetic moments inside the muffin--tin spheres are listed in 
table \ref{moments}. They agree reasonably well with earlier work. We note a 
considerable decrease of the Fe(II) moment, as compared to 
fcc Fe, while the Fe(I) moment remains almost unchanged. There is a very 
small, negative moment at the N site. \\
The decrease of the Fe(II) moment may be explained in the following way : 
Because of the strong hybridisation of the \chem{d(3z^2-r^2)} orbitals with 
N \chem{p_z} states, these \chem{d(3z^2-r^2)} states are occupied almost evenly
 in the majority and minority spin bands. Thus only 4 of the five d states are
involved in the exchange splitting of \chem{\delta^{AVG} = 1.7\, eV }. Full occupancy in the majority spin bands
and again half--filling in the minority spin bands (see Fig. \ref{dos_fe4n}) leads
to a moment of \chem{\approx 2 \mu_B}.  \\
There is a considerable charge flow from Fe to N (see table \ref{charges}), as 
may have been expected 
from simple considerations concerning the center--of--mass positions of N(p)
and Fe(3d) bands. Note that Fe(II) is more positively charged than Fe(I). From
a tight--binding analysis of the SDFT bands we find that the 4s 
partial density is reduced strongly on the Fe(II) site, and less so on the
Fe(I) site. So we may attribute the charge flow (and the resulting positive 
charges on Fe(II) and Fe(I)) to the removal of s--type states on the Fe sites
-- more or less pronounced depending on the number of and distance to the N 
ligands.

\subsection{\chem{Fe_3N}}
 
In this hexagonal lattice, all Fe sites are equivalent, as are the N sites.Again the Fe atoms form octahedra around the N sites. These octahedra share corners in such a way that N sites of adjacent ab planes are connected via the corner Fe atoms. As a consequence, each Fe site has two nearest N neighbours. The Fe-N separation is very similar to \chem{Fe_4N}, yet the N--Fe--N angle is now \chem{\approx 130^\circ} instead of \chem{180^\circ}. We note that the N sites in one specific ab plane are not connected via a N--Fe--N path. 

\begin{figure}
\epsfysize=13cm
\epsffile{ 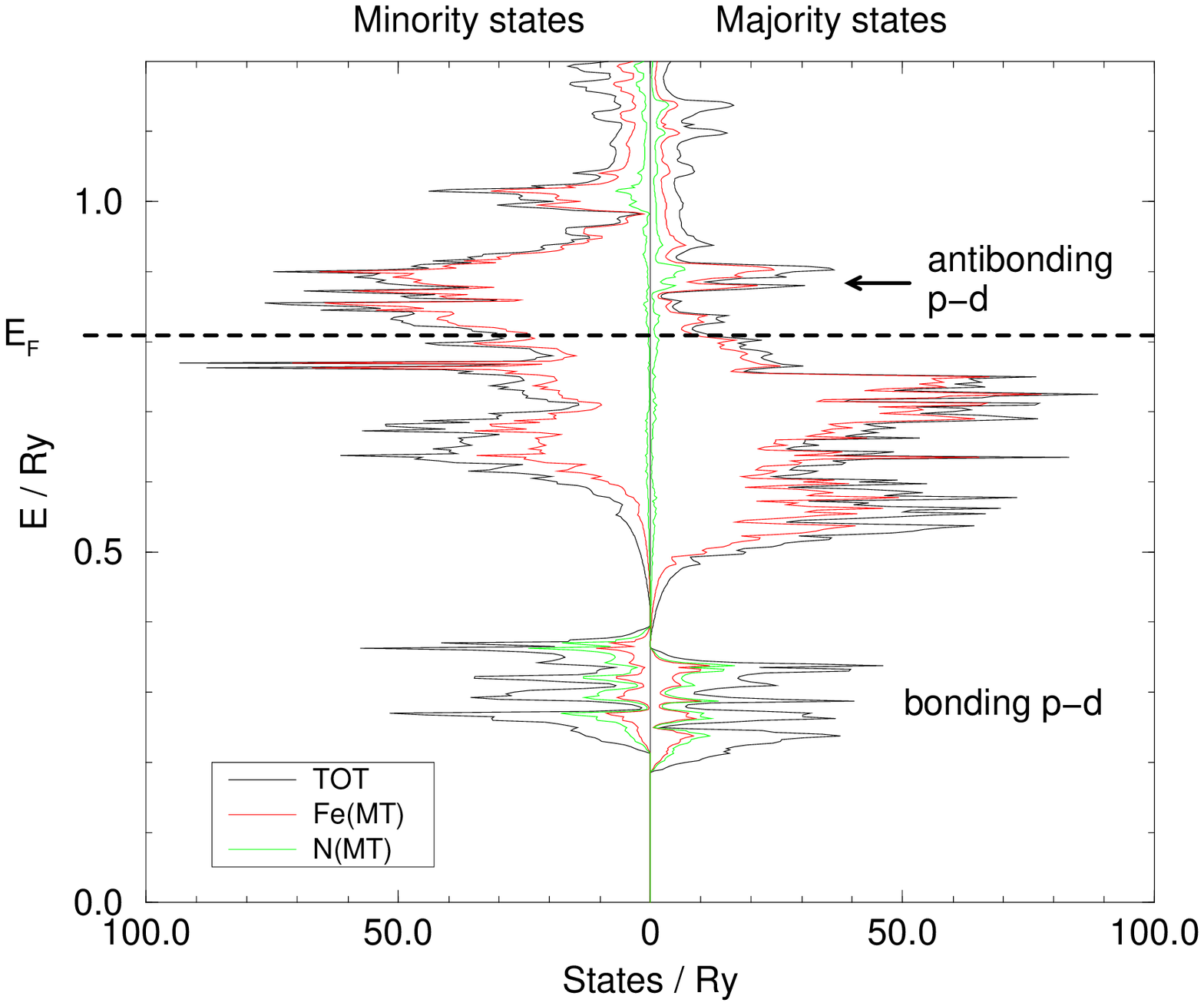}
\caption{\label{dos_fe3n} Total and muffin--tin projected  densities 
of states for \chem{Fe_3N}.}
\end{figure}

The electronic DOS curves look quite similar to the \chem{Fe_4N}
case (see Fig. \ref{dos_fe3n}). The p--d band splitting and also the exchange
splitting are very similar too, resulting in a magnetic moment on the Fe site
of ${\rm \approx 2 \mu_B}$, practically the same value as for the Fe(II) site 
in \chem{Fe_4N}. Again, this is in agreement with other studies. Each Fe site 
has two nearest N neighbours (separation 
very similar to \chem{Fe_4N}), now with an N--Fe--N angle of $\approx 
130^\circ$ instead of $180^\circ$ in \chem{Fe_4N}. We may construct on each 
Fe site a local d hybrid orbital with optimum $\sigma$ coupling to both N 
neighbours. This d hybrid is admixed into the N(p) bands in a very similar
way as is the \chem{d(3z^2-r^2)} orbital of Fe(II) into the N(p) bands of 
\chem{Fe_4N}. It is is not involved in the exchange split d--bands
(\chem{\delta^{AVG} = 1.5\, eV}).
Thus the remaining four d orbitals produce a moment of \chem{\approx 2 \mu_B}.\\
We also observe that the d minority spin bands are wider by $\approx 0.8$ eV than
the majority spin bands. This again results in negative moment 
densities in the interstitial regions. The N moment is negative and somewhat
larger than in \chem{Fe_4N}. We note that a similar negative value for N is 
also found in \chem{\zeta - Fe_2N} (see table \ref{moments}). From Fig. \ref{dos_fe3n}
we realize that in the N(p) bands the ratio of p to d partial densities is
larger for the minority bands (also observed for \chem{\zeta - Fe_2N} and to a lesser 
extend for \chem{Fe_4N}). We attribute this effect to the larger energy 
difference between the centers of N(p) and Fe(d) bands in the minority spin case. This reduces p--d hybridisation there and enhances the p--type component 
over the majority spin case. \\
The charge density distribution is also very similar to \chem{Fe_4N}  --
strong p--d bonding charge along the Fe--N bond and weaker Fe--Fe bonding. The
resulting ionic charge of N has the same value as for \chem{Fe_4N}, the Fe 
atoms are positively charged with values slightly larger than for the Fe(II) site 
of \chem{Fe_4N}. Again, the charge flow can be attributed to the removal of 
s--type states on the Fe sites. \\

\subsection{\chem{\zeta - Fe_2N}}

This lattice type may be seen as a slightly distorted hexagonal structure. Again, there is only one type of Fe sites and one type of N sites. Like in \chem{Fe_3N} the N centered octahedra share corners when the N atoms are situated in adjacent ab planes of the hexagonal lattice. Unlike \chem{Fe_3N}, the N centered octahedra of one ab plane now share edges. 

\begin{figure}
\epsfysize=13cm
\epsffile{ 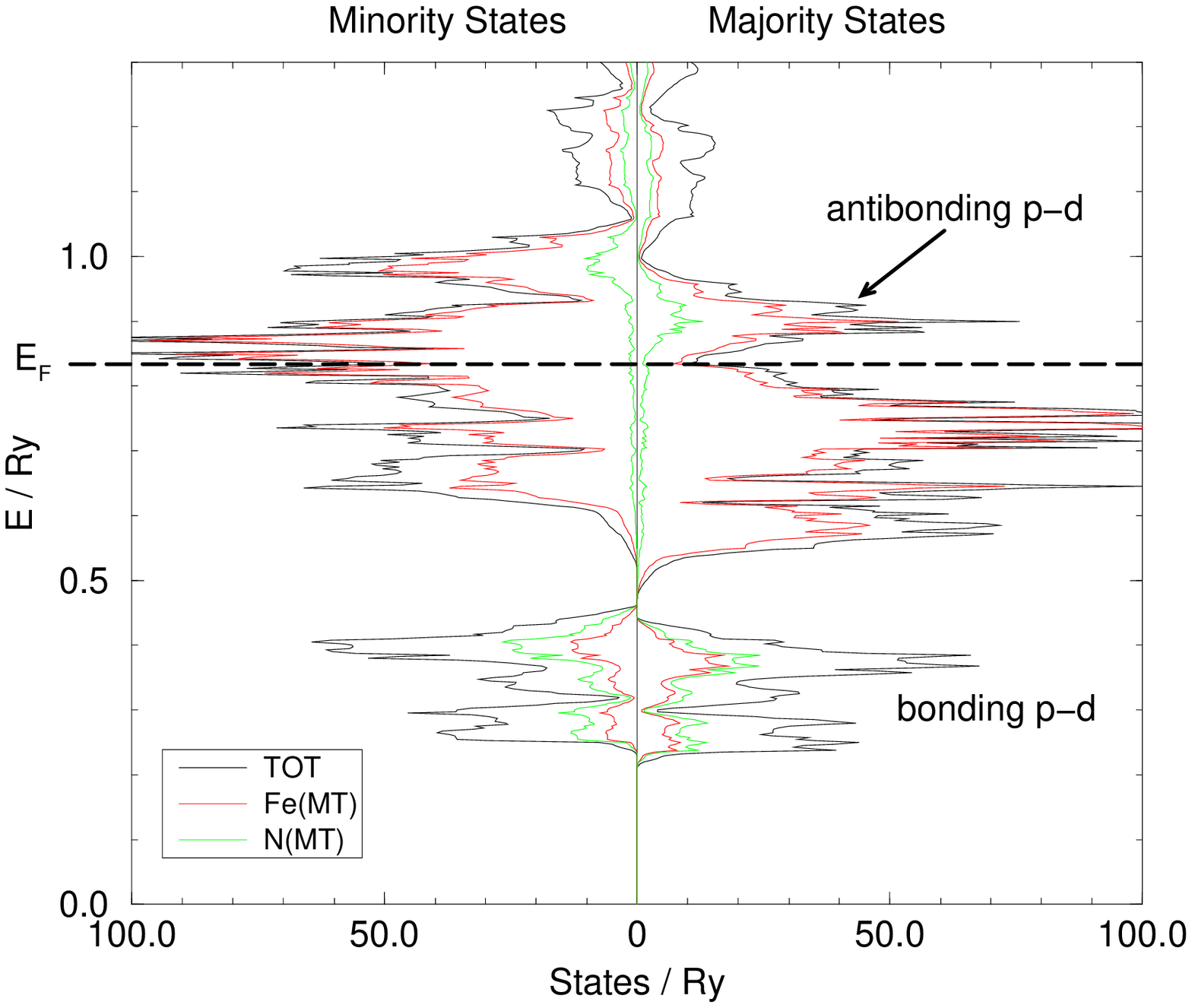}
\caption{\label{dos_fe2n} Total and muffin--tin projected densities 
of states for \chem{Fe_2N}.}
\end{figure}

As a consequence, each Fe has 3 nearest N
neighbours. If we try to construct a local basis with optimum $\sigma$ 
coupling of Fe(d) hybrids to the N(p) orbitals, we require two hybrids.
These are then excluded from the exchange split d--bands. Assuming again fully
occupied majority and half--filled minority bands, we arrive at a moment of 
1.5 \chem{\mu_B}. Looking at Fig. \ref{dos_fe2n} we find this model reasonably
well justified, although there are some deviations to the situation of 
\chem{Fe_3N} : The exchange splitting is reduced down to 
\chem{\delta^{AVG} \approx 1.2 \, eV}
, the anti--bonding p--d bands
overlap slightly with the top of the pure d--bands and also the width of the 
minority spin 
bands (\chem{\approx 4.3 \, eV}) is approaching the value of the majority spin
 bands (\chem{\approx 4.0 \, eV}). \\
The charge density maps again indicate the bonding charge along the Fe(d) --
N(p) bond. The N charge has decreased slightly in absolute value, indicating 
that the reservoir of s--type charge on the Fe atoms is almost exhausted. \\

\subsection{\chem{FeN}, rock salt structure}

Here, all adjacent N centered octahedra share edges. There are now six nearest neighbours of 
N to each Fe. The two d orbitals which exhibit $\sigma$ 
coupling to N(p) are the \chem{E_g} orbitals \chem{d(3z^2-r^2)} and 
\chem{d(x^2-y^2)}. Again, we may expect a moment of 1.5 \chem{\mu_B}, but we 
realize that the overlap of the anti--bonding \chem{E_g} bands with the 
\chem{T_{2g}} bands is quite significant. Also, the exchange splitting is 
further reduced down to \chem{\delta^{AVG} \approx 1.0 \, eV} , and, finally the excess width of the minority bands is 
smaller again. The negative charge at the N site is about the same as in 
\chem{\zeta - Fe_2N},
but now all s--type charge density flow is exhausted, and a reduction of the
d--type density is observed at Fe. Most of these effects help to reduce the 
Fe moment to \chem{\approx 1.1 \mu_B}. Now the N moment is also positive. We 
attribute this effect to the above mentioned band overlap of the anti--bonding
\chem{E_g}--p bands and the \chem{T_{2g}} bands. For the majority spin case,
this leads to considerable extra majority spin density at N. There are still
interstitial regions with negative moment density, these are now restricted 
to the tetrahedral interstitial sites, as all octahedral sites are occupied by
N.\\

\section{Summary}

There are clear trends seen in our study of \chem{FeN_x} with close packed 
Fe lattices: \\
A) Bonding : N occupies octahedral sites. A strong covalent p--d bonding takes
place in those octahedra, concomitant with charge transfer to N. These extra
bonds clearly add stability to the Fe lattice. At low x values, the charge
flow is predominantly fed by the s--type charge of the Fe sites, and the N 
excess charge is as high as -1.4. Values of that magnitude have been found in
other compounds involving transition metals, such as \chem{CaTaN_2} 
\cite{Smolinski}. In 
nitrides with divalent simple metals or with noble metals \cite{Hahn}, an 
even larger nitrogen charge of up to \chem{Z_N} = -1.8 has been found. \\
With increasing N content, the N charge is more and more reduced down to a 
value of 
\chem{Z_N \approx -1.1} for hypothetical FeN of rock salt structure. \\
B) Magnetism : There holds a rather simple rule through the series of Fe nitrides. The 
starting point is fcc Fe in the high spin state, carrying a moment of 
\chem{\approx 2.5 \mu_B}. The five majority spin d--bands are completely filled
(the case of a 'strong' ferromagnet), while the five minority states are half--filled, leading to a moment of 
\chem{5 \cdot \frac{1}{2} \mu_B}. \\
In cubic \chem{Fe_4N}, there are two inequivalent Fe sites. Fe(I) has 
the same nearest neighbour shell as in fcc iron, and 
keeps its moment. Fe(II) has two N nearest neighbours. 
Of the five
d--orbitals, the \chem{d(3z^2-r^2)} shows strong $\sigma$--type coupling to 
N \chem{p_z}
(leading to the strong covalent bond discussed above). This orbital does not 
participate in the exchange splitting of the other d bands. Again, we find a 'strong' ferromagnet case, and the minority bands are half-filled. As a consequence, the
magnetic moment of the Fe(II) site is reduced to \chem{4 \cdot \frac{1}{2} 
\mu_B}. \\
In hcp \chem{Fe_3N}, all the Fe sites are equivalent, with two nearest 
neighbours of N. Using a similar construction of the covalent bonds we again 
arrive at 4 exchange split bands per Fe and a moment of 2 \chem{\mu_B}. \\
In \chem{\zeta - Fe_2N}, each of the equivalent
Fe sites has three nearest neighbours of N. Now two d orbitals per Fe are 
involved in the covalent bonds, and the three remaining exchange split d bands 
per Fe yield a moment of \chem{\approx 1.5 \mu_B}. \\
Even for hypothetical FeN with rock salt structure, this scheme holds 
approximately. Now there are six nearest nitrogen neighbours and the two 
\chem{E_g} orbitals participate in the covalent bonds. Again a moment of 
\chem{\approx 1.5 \mu_B} is expected from the \chem{T_{2g}}--type Fe d bands.
Yet the reduced exchange splitting and other effects 
reduce the moment to 1.1 \chem{\mu_B}. \\
Another notable effect is the large excess of the minority spin bandwidth 
over that of the majority spin bands, caused by the radially expanded d--orbitals of the half filled minority spin bands.
One of the consequences is the appearance of negative--moment
density in the octahedral and tetrahedral interstitial regions. \\

\acknowledgments
This work was supported by the DFG. Some of the calculations were performed
on the Cray T3E of the Höchstleistungsrechenzentrum Jülich and the IBM SP2
of the GMD.

\newpage
\newpage

\newpage

\newpage

\newpage
\begin{figure}
\epsfysize=8 cm
\hspace{1.5cm}
\epsffile{ 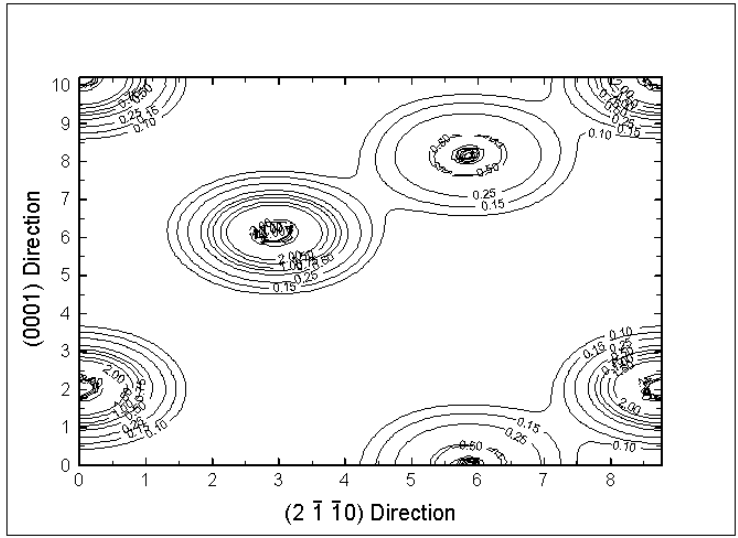}
\caption{\label{charge_fe3n} Valence charge density of \chem{Fe_3N} in the 
(03$\bar{3}$0) plane. The N atoms are lying in the plane, while the Fe
atoms are centered slightly above.}
\end{figure}

\begin{figure}
\epsfysize=8 cm
\hspace{1.5cm}
\epsffile{ 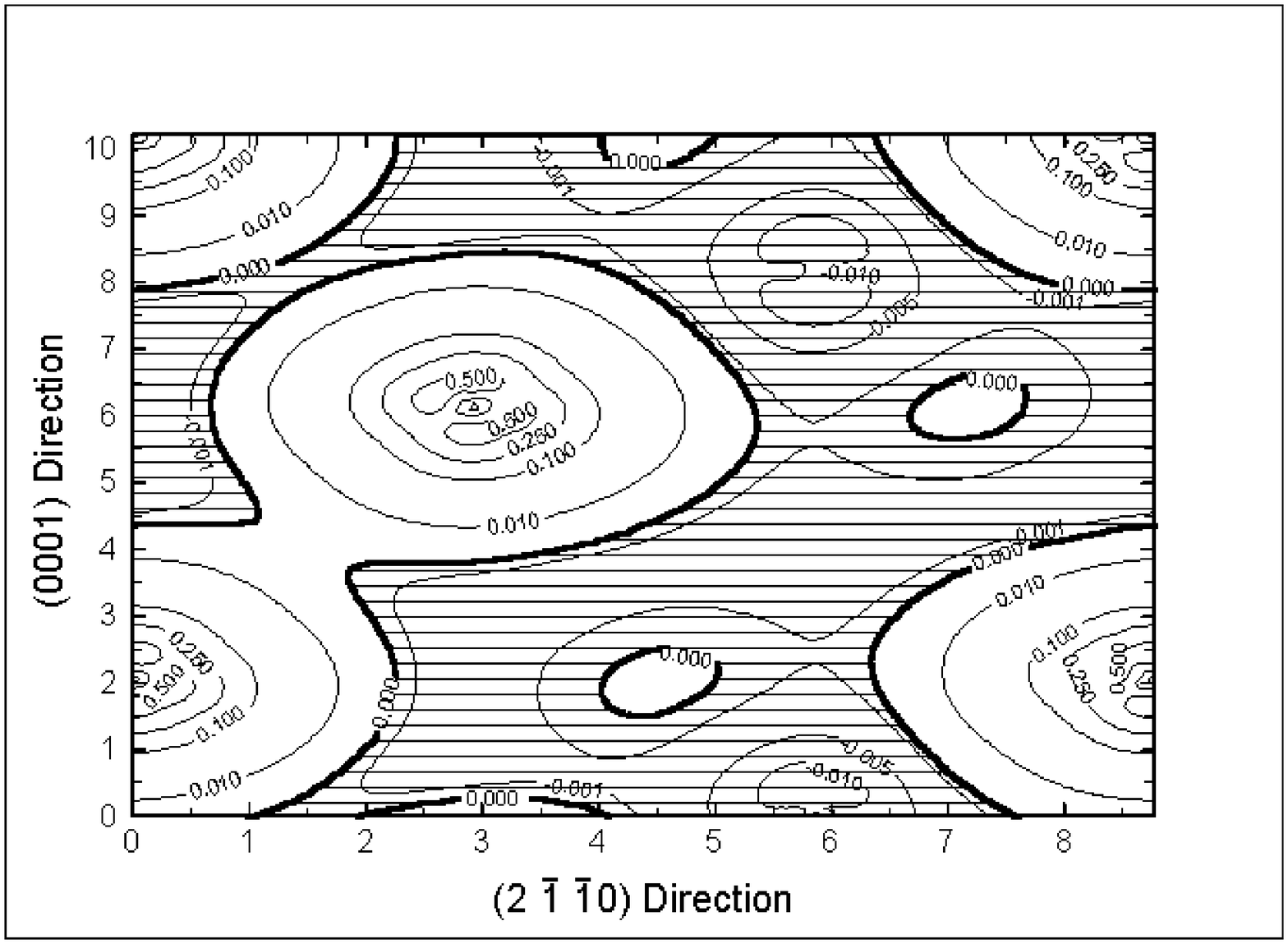}
\caption{\label{spin_fe3n} Spin density of \chem{Fe_3N} in the 
(03$\bar{3}$0) plane.}
\end{figure}
\newpage


\newpage

\begin{figure}
\epsfysize=8 cm
\hspace{1.5cm}
\epsffile{ 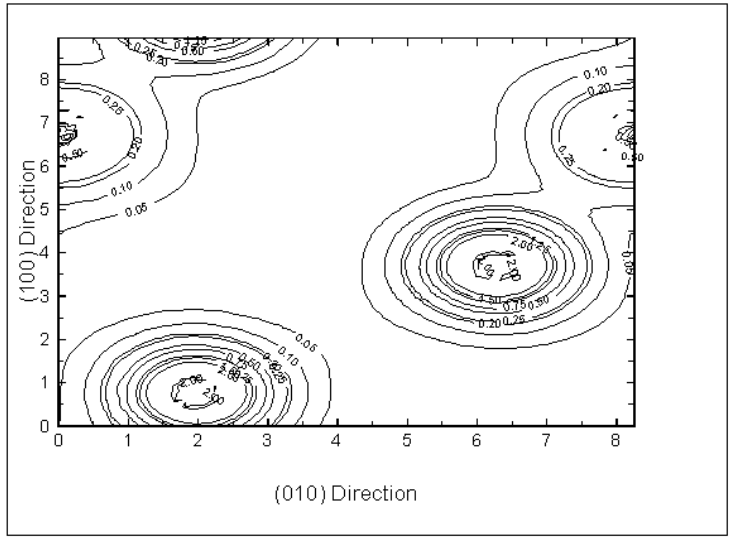}
\caption{\label{charge_fe2n} Valence charge density of \chem{Fe_2N}. 
The Fe atoms are lying in the plane, but 
the N atoms are centered slightly above the plane.}
\end{figure}

\begin{figure}
\epsfysize=8 cm
\hspace{1.5cm}
\epsffile{ 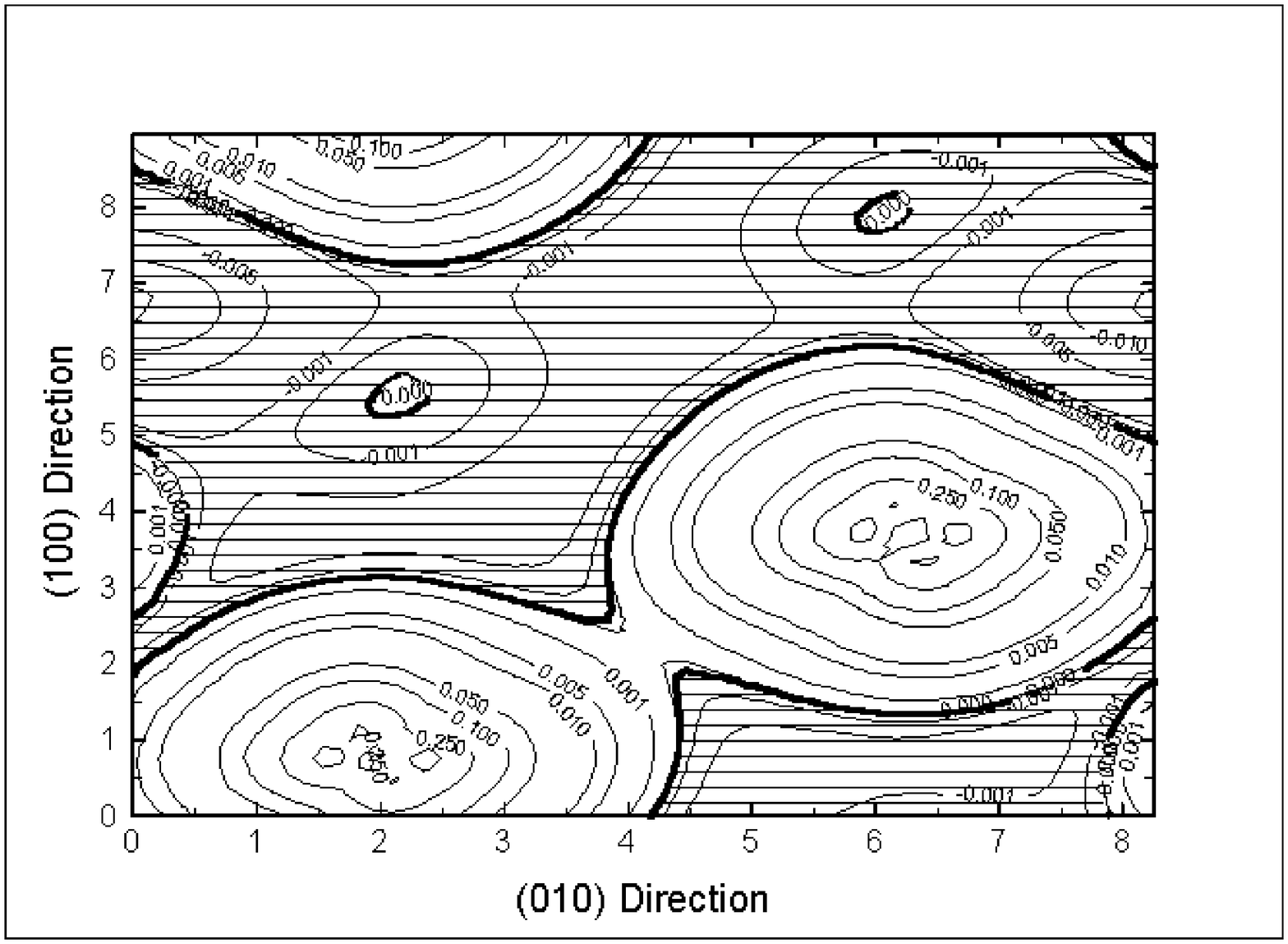}
\caption{\label{spin_fe2n} Spin density of \chem{Fe_2N}. }
\end{figure}
\newpage



\begin{figure}
\epsfysize=13cm
\epsffile{ 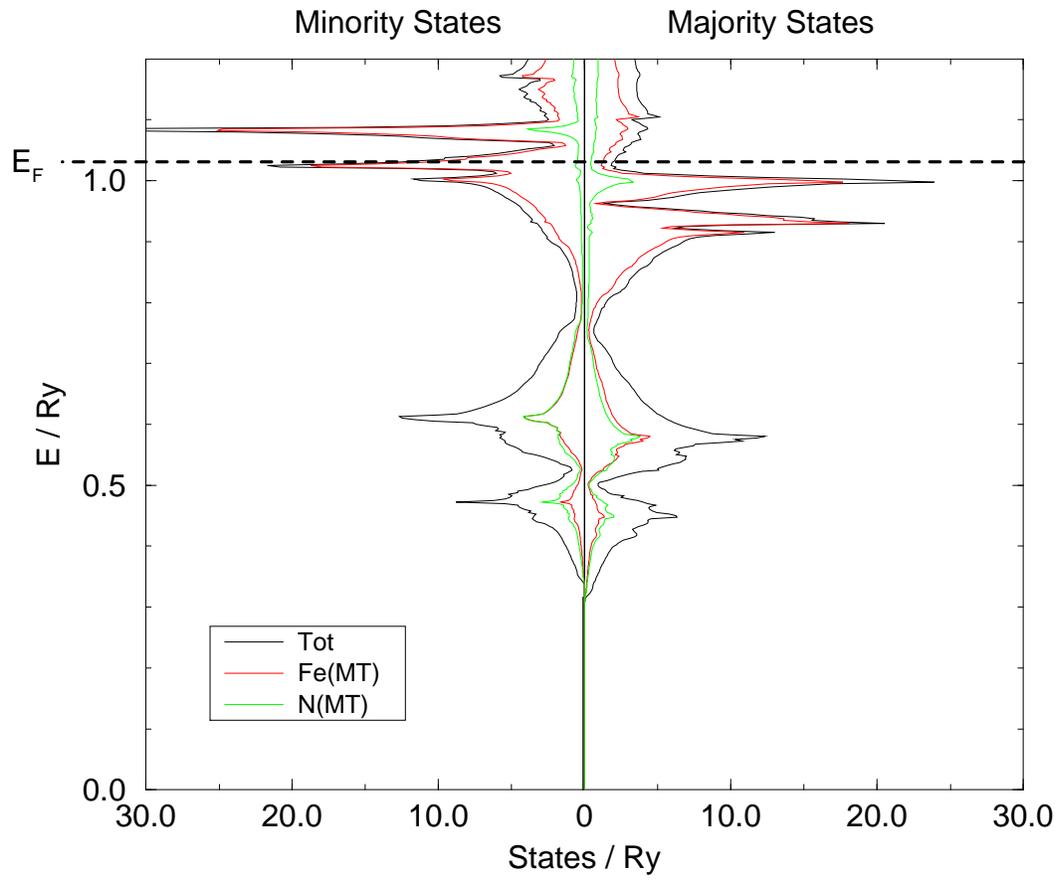}
\caption{\label{dos_fen} Total and muffin--tin partial densities 
of states for \chem{FeN} in the rocksalt structure.}
\end{figure}
\newpage
\begin{figure}
\epsfysize=8 cm
\hspace{1.5cm}
\epsffile{ 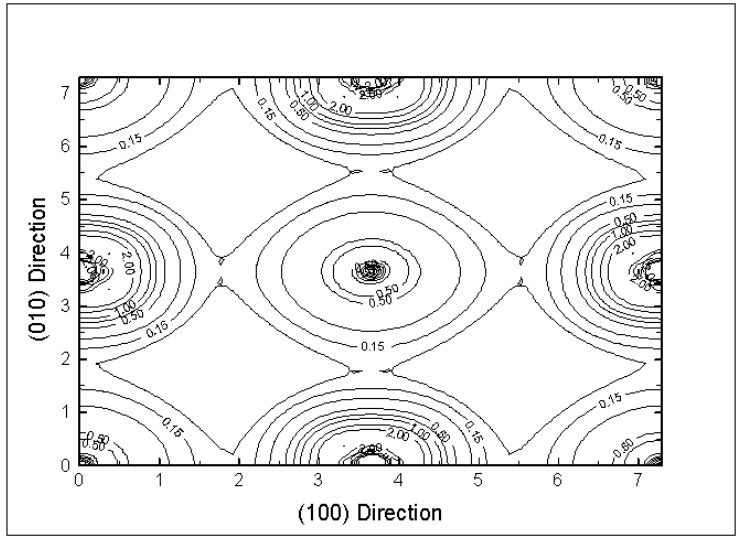}
\caption{\label{charge_fen} Valence charge density of \chem{FeN(NaCl)} in the (100) 
plane.}
\end{figure}

\begin{figure}
\epsfysize=8 cm
\hspace{1.5cm}
\epsffile{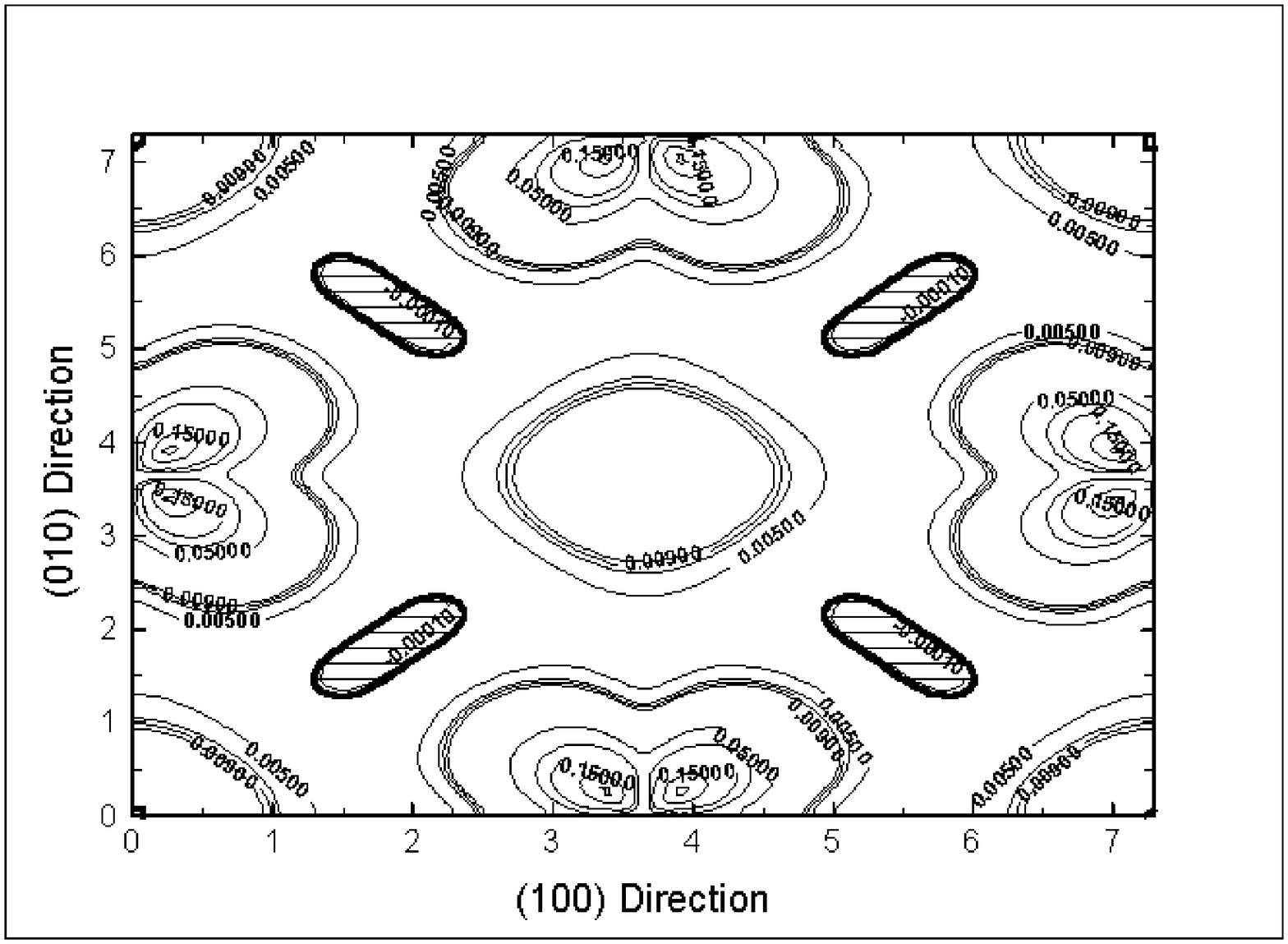}
\caption{\label{spin_fen} Spin density of \chem{FeN(NaCl)} in the (100) 
plane.}
\end{figure}


\begin{table}
\caption{\label{structures} Structures and lattice parameters of the various 
iron nitrides. All lengths are given in atomic units (a.u.).}

\begin{tabular}{cccccccc}
Material & space group & a / a.u. & b / a.u. & c / a.u. & Fe(I) & Fe(II) & N \\
\tableline
\chem{Fe_4N} & \chem{Pm\bar{3}m} & 7.1707 & 7.1707 & 7.1707 & 3d & 1b & 1a \\
             &                   &        &        &        & 
(\half,0,0) & (\half,\half,\half) & (0,0,0) \\
\tableline
\chem{Fe_3N} & \chem{P6_322} & 8.8694 & 8.8694 & 8.2552 & -- & 6g & 2c \\
             &               &        &        &        &    & (0,0.325,0) & $(\frac{1}{3},\frac{2}{3},\frac{1}{4}) $ \\
\tableline 
\chem{\zeta - Fe_2N} & \chem{Pbcn} & 8.3881 & 10.4751 & 9.1548 & - & 8d & 4c \\
             &             &        &         &        & -- & (0.251,0.128,0.008) & (0,0.864,0.25) \\
\tableline
\chem{FeN}   & \chem{Fm\bar{3}m} & 7.372 & 7.372 & 7.372 & -- & 4a & 4b \\
             &                    &      &       &       & -- & (0,0,0) & $(\frac{1}{2},\frac{1}{2},
\frac{1}{2})$ \\ 
\end{tabular}
\end{table}

\begin{table}
\caption{\label{charges} Ionic charges as determined by the method of Bader.
Also shown are charges of previous work, derived from the (overlapping) 
muffin--tin occupation numbers.}
\begin{tabular}{lccc}
Material & \multicolumn{3}{c}{Charges} \\
\tableline
 &Fe(I) & Fe(II) & N  \\
\tableline \tableline
\chem{Fe_4N} & +0.2 & +0.4 & -1.4 \\
\chem{Fe_4N}\tablenote{Ref. \cite{Matar1}} & +1.04 & +0.51 & +0.02 \\
\tableline	
\chem{Fe_3N} & -- & +0.47 & -1.4 \\
\chem{Fe_3N}\tablenote{Ref. \cite{Matar2}} & -- & -0.02 & +0.95 \\
\tableline
\chem{\zeta - Fe_2N} & -- & +0.6  & -1.2 \\
\chem{\zeta - Fe_2N}\tablenote{Ref. \cite{Matar3}} & --& -0.23 & +0.89 \\
\tableline 
\chem{FeN} & -- & +1.1 & -1.1  \\
\end{tabular}
\end{table}

\begin{table}
\caption{\label{moments} Muffin--tin magnetic moments of the various materials.
We also give the results of previous work. All magnetic moments are given in units of 
\chem{\mu_B}.}
\begin{tabular}{lcccc}
Material & \multicolumn{3}{c}{Muffin--tin moment} \\
\tableline
 &Fe(I) & Fe(II)  & N \\
\tableline \tableline
\chem{Fe_4N} & 2.73 & 1.97 & -0.005 \\
\chem{Fe_4N}\tablenote{Ref. \cite{Matar1}} & 2.98 & 1.79 & 0.02 \\
\tableline
\chem{Fe_3N} & -- & 1.99 & -0.06 \\
\chem{Fe_3N}\tablenote{Ref. \cite{Matar2}} & -- & 1.96 & -0.05 \\
\tableline
\chem{\zeta - Fe_2N} & -- & 1.43 & -0.06 \\
\chem{\zeta - Fe_2N}\tablenote{Ref.  \cite{Matar3}} & -- &  1.49 & -0.07 \\
\tableline 
\chem{FeN} & -- & 1.15 & 0.08 \\
\end{tabular}
\end{table}


\end{document}